\documentclass[pre,twocolumn,showpacs,preprintnumbers,amsmath,amssymb,floatfix]{revtex4}
\usepackage{bm,graphicx}
\begin{document}
 

\title{Analytical characterization of adhering vesicles}

\author{C. Tordeux}
\author{J.-B. Fournier}
\affiliation{Laboratoire de Physico-Chimie Th\'eorique,
E.\,S.\,P.\,C.\,I., 10 rue Vauquelin, F-75231 Paris cedex 05, France}
\author{P. Galatola}
\affiliation{LBHP, Universit\'e Paris 7---Denis Diderot, Case 7056, 2
place Jussieu, F-75251 Paris cedex 05, France}

\date{\today}
\begin{abstract}
We characterize vesicle adhesion onto homogeneous substrates by means of a
perturbative expansion around the infinite adhesion limit, where curvature
elasticity effects are absent. At first order in curvature elasticity, we
determine analytically various global physical quantities associated with
adhering vesicles: height, adhesion radius, etc. Our results are valid for
adhesion energies above a certain threshold, that we determine numerically.
We discuss the haptotactic force acting on a vesicle in the limit of weak
adhesion gradients. We also propose novel methods for measuring adhesion
energies and we suggest a possible way of determining the size of suboptical
vesicles using controlled adhesion gradients.
\end{abstract}

\pacs{87.16.Dg, 68.35.Np, 68.03.Cd}

\maketitle


\section{Introduction}

When phospholipids are dissolved in an aqueous solution, almost all the
molecules condensate into bilayers. Lipid bilayers are formed by two
contacting monolayers of opposite orientation, in which the hydrophilic heads
of the molecules are located at the sides of the structure, the hydrophobic
tails being shielded from contact with water~\cite{israelachvili_book}.  As
there is a prohibitive energy cost associated with their free borders, these
bilayers form closed objects, which are called vesicles.  For some biological
studies, vesicles are used as models of the membrane of living
cells~\cite{sternberg87}. They also have applications as encapsulation vectors
for drug delivery~\cite{guedeau-boudeville95}.  Their efficiency as drug
delivery vectors is linked to their permeability, which can be affected by
adhesion phenomena~\cite{bernard00}. Vesicle adhesion on a solid substrate,
followed by its rupture and fusion, also provides a simple technique for
obtaining supported membranes~\cite{keller00} that can be used for the design
of biosensors~\cite{sackmann96}.

Adhesion phenomena between a lipid bilayer and a substrate can be divided into
two categories: i) specific adhesion between a particular host protein and a
receptor on the substrate~\cite{bruinsma00}; this kind of adhesion generally
implies a process of molecular recognition between a receptor and a ligand,
and is common in biological systems.  ii) non-specific adhesion between the
membrane's lipids and the substrate, mediated by universal interactions, e.g.,
van der Waals forces. Here, we focus on non-specific adhesion, which can be
described by an adhesion potential $W$ that represents the free energy gain
per unit area of contact. Typical values of $W$ range from
$10^{-4}\,$mJ/m$^2$ to $1$\,mJ/m$^2$~\cite{seifert97}. Note that the
description of adhesion using a contact potential is approximate,
because van der Waals forces are actually long-ranged and because
membranes may fluctuate in the vicinity of the substrate: adhering
vesicles actually never strictly come into contact with their substrate.
Membrane--substrate separations range from $1$\,nm for the strongest
values of $W$~\cite{sackmann96}, to about $50$\,nm for the weakest
adhesions~\cite{radler95}. The highest values of $W$ tend to produce
vesicle rupture during the adhesion process~\cite{keller00}, owing to a
strong tension induced in the membrane~\cite{seifert97}.

To determine the shape and free energy of an adhering vesicle, one must take
into account the competition between: i) the adhesion energy gain, ii) the
constraints on the total membrane area~$A$ and the total enclosed volume~$V$,
and iii) the free energy cost associated with the curvature elasticity of the
membrane. The latter is described by a free energy density proportional to the
square of the local mean curvature~\cite{helfrich73}. For lipids, the
corresponding bending rigidity $\kappa$ is of the order of
$10^{-19}\,\mathrm{J}\simeq25\,k_\mathrm{B}T$ at room
temperature~\cite{lee01}.  Refined vesicle models take into account a
constraint on the difference between the areas of the two
monolayers~\cite{svetina82}, or an elasticity associated with
it~\cite{seifert92}. Physically, this arises from the fact that lipids are
not significantly exchanged between the two monolayers during typical
experimental times. It is not known at the present time whether this
constraint is significant for adhering vesicles: to simplify, we shall
disregard it in our approach. 

The shapes of axisymmetric adhering vesicles can be determined by functional
minimization~\cite{seifert90,seifert97}. However, due to non-linearities in
the equilibrium equations, exact solutions can only be determined numerically.
In the asymptotic case of infinitely strong adhesion, $W\to\infty$ (or
equivalently $\kappa\to0$), the problem is easily solved
analytically~\cite{seifert90}: the equilibrium shapes are spherical caps,
whose features are dictated by the geometrical constraints only. In this paper
we characterize the adhesion of vesicles in the case of strong but finite
adhesion, by extracting analytical corrections with respect to the infinite
adhesion case.  We determine analytically the first-order corrections to
various physical observables and we discuss their limit of validity by a
direct comparison with exact numerical results.

\begin{figure}
\centerline{\includegraphics[width=7.5cm]{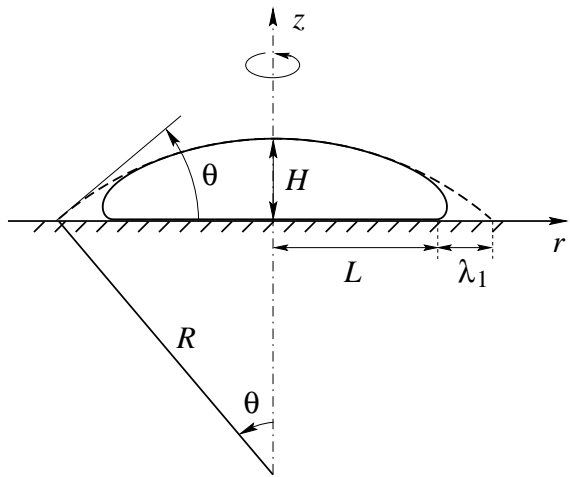}}
\caption{Definition of the global observables associated with an
axisymmetric adhering vesicle. The vesicle's shape, which was calculated
numerically, corresponds to a rather deflated situation in which the
``contact angle region'' is broad and not well-defined. When adhesion is
stronger, the vesicle's shape resembles a spherical cap (dashed line),
with a strongly curved rim at the foot of the ``contact angle'' (no
discontinuity of the membrane's normal).}
\label{schema_general}
\end{figure}

The first-order corrections with respect to the infinite adhesion limit
originate from the existence of a strongly curved region at the border
of the adhesion disc (see Fig.~\ref{schema_general}). We shall refer to
this region as the ``contact angle region", by analogy with wetting
phenomena~\cite{deGennes85}. The shape of this region has been
determined in Refs.~\cite{servuss89,guttenberg00} using an open membrane
description, i.e., no volume constraint and an externally imposed
tension acting along a fixed direction mimicking the asymptotic contact
angle.  Imposing explicitly the volume constraint, we recover the same
shape for the contact angle region. The novelty of our approach resides
in the analytical description of the various observables associated with
the adhering vesicle (height, radius of adhesion, $\ldots$).

Standard measurements of adhesion potentials~$W$ are based on the
determination of the shape of the contact angle region, e.g., by Reflection
Interference Contact Microscopy (RICM). Indeed, the radial curvature~$c$ of a
detaching membrane yields $W$ through the equilibrium relation
$c=\sqrt{2W/\kappa}$~\cite{seifert90,landau_book_elasticity,rosso99}. In
practice, it is difficult to precisely measure $c$, and it is more efficient
to fit the contact angle region using RICM~\cite{nardi98,guttenberg00}.
Available models rest however on linearized equations for contact angles close
to $\pi$~\cite{guttenberg00}.  Our non-linear analysis allows not only to fit
the contact-angle region and determine contact potentials even for contact
angles far from~$\pi$, but also provides new means of determining $W$ by
measuring the various global observables. 

Our paper is organized as follows: In Sec.~\ref{presentgen}, we
introduce the model used to describe the elasticity of vesicles and
their adhesion onto homogeneous substrates. We also define various
global observables relevant to the adhesion geometry.
Section~\ref{raccord} contains the results of our analytical
calculations: in Sec.~\ref{Winfini}, we recall the asymptotic limit of
infinite adhesion; in Sec.~\ref{finiteadh}, we recall the general
equations describing the equilibrium shapes of adhering axisymmetric
vesicles; in Sec.~\ref{CAshape} we calculate the shape of the contact
angle region at first-order in $\sqrt{\kappa/(WA)}$; in
Sec.~\ref{extrapol}, we determine the contact angle extrapolation
length~\cite{nardi98,guttenberg00}; in Sec.~\ref{order1}, we determine
the first-order expansions, in power series of $\sqrt{\kappa/(WA)}$, of
the various global observables associated with the vesicle's shape; in
Sec.~\ref{freeen}, we calculate at first-order the free energy of
adhering vesicles and we discuss haptotaxis (motion induced by an
adhesion gradient)~\cite{cantat00}. In Sec.~\ref{ordre1lim}, we
determine \textit{numerically} the global observables and we discuss the
range of validity of the corresponding first-order expansions.  Finally,
in Sec.~\ref{conclusion}, we summarize our results and we discuss some
possible applications, including new methods for measuring $W$.

\section{Description of adhering vesicles}\label{presentgen}

In most experimental situations, although vesicles are slightly permeable to
water, their volume~$V$ is strongly fixed by the osmotic pressure of the
various solubilized ions to which the membrane is
impermeable~\cite{seifert97}. We shall suppose that this volume constraint
remains satisfied for adhering vesicles. The area~$A$ of vesicles is also
fixed to a high accuracy: solubilized lipids are almost inexistent
and the area-stretching modulus~$k_s$, which is of the order of
$100$\,mJ/m$^2\gg W$, cannot significantly affect the area
constraint~\cite{seifert97}. It is traditional to introduce a dimensionless
parameter~$v$, the reduced volume, defined by
\begin{equation}\label{reducedv}
v = \frac{V}{\frac{4}{3}\pi\left(A/4\pi\right)^{3/2}}.
\end{equation}
This quantity $0<v\le1$ describes how much the vesicle is deflated with
respect to a sphere ($v=1$). Due to the constraints, it is fixed.

$V$ and $A$ being fixed, the free energy of an adhering
vesicle is given by
\begin{equation}\label{freeneconst}
F = - W A_\mathrm{adh} 
+\oint dA\,\, \frac{1}{2} \kappa(c_1 + c_2)^2,
\end{equation}
where  $c_1$ and $c_2$ are the two local principal curvatures of the membrane,
$\kappa$ is the Helfrich bending constant~\cite{helfrich73}, $A_\mathrm{adh}$
is the area of contact between the vesicle and the substrate, and $W$ is the
contact potential. As discussed in the Introduction, we simply model the
adhesion by an energy proportional to the contact area. In principle, $F$
should also contain a Gaussian curvature term $\bar\kappa\,c_1c_2$, however
we discard it since its integral over the membrane is constant for a given
vesicle topology, according to the Gauss-Bonnet theorem~\cite{struik_book}.
Therefore, there are only two dimensionless parameters in the problem: $v$ and
$\kappa/(W A)$.

In the whole paper, we shall restrict ourselves to axisymmetric vesicle
shapes.  We define the following global observables (see
Fig.~\ref{schema_general}): we call $H$ the height of the vesicle measured on
the revolution axis, and $L$ the radius of the adhesion disc. The adhering
area is thus $A_\mathrm{adh}=\pi L^2$.  In the regime of strong adhesion,
vesicles almost take the shape of a spherical cap.  In order to precisely
define a ``contact angle'' even in the case of weaker adhesion, we introduce
the sphere which is osculatory to the membrane at the point intersecting the
revolution axis. We call $R$ its radius and $\theta$ the angle at which it
intersects the substrate (see Fig.~\ref{schema_general}).  Finally, we 
define the extrapolation length $\lambda_1$ as the distance between the point
where the vesicle detaches from the substrate and the intersection between the
osculatory sphere and the substrate.

We shall denote throughout by the index zero all the quantities
referring to the limit $W\to\infty$, where the vesicle exactly takes the
shape of a spherical cap. Therefore  $H_0$, $R_0$, $\theta_0$ and $L_0$ are
the height, radius, contact angle, and adhesion radius of the corresponding
spherical cap.

\section{Analytical results} \label{raccord}

Strong adhesion corresponds to the situation where the adhesion 
energy gain is very large compared to the elastic energy of the vesicle.
Since the energy of freely floating vesicles is of order
$\kappa$~\cite{seifert97}, even for deflated vesicles, this condition
can be expressed as
\begin{equation}\label{strgadhcond}
W A \gg \kappa.
\end{equation}
It corresponds, for a given vesicle, to strong enough contact
potentials~$W$, or, for a given $W$, to large enough vesicles. In this
situation, elasticity can be treated as a first-order correction with
respect to the asymptotic limit of infinite adhesion. We shall therefore
first review the limit $W\to\infty$~\cite{seifert90}.

\subsection{Infinite contact potential $W$}
\label{Winfini}

In this case, adhesion is the only relevant contribution to the free energy of
the system, and $v$ is the only dimensionless parameter of the problem.
Taking into account the two geometrical constraints, and formally setting
$\kappa=0$, the shape of the adhering vesicle is deduced from the minimization
of the following functional:
\begin{equation}\label{Fkappa=0}
F_0^\star = - W A_\mathrm{adh}^{0} + \Sigma_0 A + P_0 V.
\end{equation}
$\Sigma_0$ is the Lagrange multiplier associated with the area
constraint and $P_0$ is the Lagrange multiplier associated with the
volume constraint. Equation~(\ref{Fkappa=0}) can be rewritten as
\begin{equation}\label{Fkappa2=0}
F_0^\star = \left(\Sigma_0 - W\right) A_\mathrm{adh}^{0} 
+ \Sigma_0 \left(
                                    A - A_\mathrm{adh}^0
                                                              \right) 
            + P_0 V.
\end{equation}
This functional is identical to that of a liquid droplet wetting a flat
substrate, with the correspondence
$\Sigma_0-W\to\gamma_\mathrm{SL}-\gamma_\mathrm{SV}$,
$\Sigma_0\to\gamma_\mathrm{LV}$ and $P_0\to-\Delta P$, in which
$\gamma_\mathrm{SL}$, $\gamma_\mathrm{SV}$, $\gamma_\mathrm{LV}$ have
their usual meaning and $\Delta P$ is the drop's excess
pressure~\cite{deGennes85}.  This implies that infinitely strongly
adhering vesicles and liquid droplets have the same ensemble of
equilibrium shapes, although they are described by different sets of
physical parameters. Consequently, the equilibrium shapes in the
asymptotic limit $W\to\infty$ are spherical caps. 

The major physical difference with the case of liquid droplets is that
the contact angles are not fixed by surface tensions, but rather by the
geometrical constraints acting on the vesicles.  The relation between
the contact angle~$\theta_0$ and the reduced volume $v$
[Eq.~(\ref{reducedv})] can easily be deduced from simple
geometry~\cite{seifert97}: 
\begin{equation}\label{vdetheta}
v = \frac {8 - 9 \cos \theta_0 +\cos3\theta_0 }
		{2\left( 2 - 2 \cos\theta_0 + \sin^2\theta_0 \right)^{3/2}}.
\end{equation}
As for the two Lagrange multipliers, they can easily be found by using the
analogy with wetting droplets: the Young relation
$\gamma_\mathrm{LV}\cos\theta+\gamma_\mathrm{SL}=\gamma_\mathrm{SV}$ yields
\begin{equation}\label{youngrelat}
\Sigma_0 = \frac{W}{1 + \cos\theta_0},
\end{equation}
and the Laplace law $\Delta P=2\gamma_\mathrm{LV}/R_0$ yields
\begin{equation}
\label{Laplacerelat}
P_0 = - \frac{2 \Sigma_0}{R_0}
= - \frac{2 W \sin\theta_0}{L_0 \left(1 + \cos\theta_0 \right)}.
\end{equation}

\begin{figure}
\centerline{\includegraphics[width=8cm]{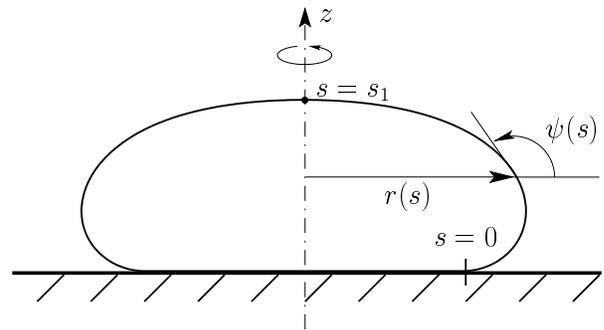}}
\caption{Definition of the parameters used in the determination of the
equilibrium shape of an adhering vesicle.}
\label{param}
\end{figure}

\subsection{The equations describing finite adhesion}\label{finiteadh}

Let us now consider the case of a finite contact potential~$W$. The
equilibrium shapes are those minimizing the sum of the bending free energy and
the adhesion free energy, subject to the area and volume constraints.
Considering axisymmetric shapes, we parameterize their contour by the tangent
angle~$\psi(s)$, where $s\in[0,s_1]$ is the arc-length (see Fig.~\ref{param}),
such that at $s=0$ the membrane leaves the substrate and at $s=s_1$ it attains
the revolution axis. Although $\psi(s)$ alone is sufficient to describe the
vesicle's shape, it is more convenient to also introduce the distance $r(s)$
to the revolution axis~\cite{seifert91}. In the following, we shall denote by
a dot derivation with respect to $s$.  The two principal curvatures are
$c_1=\dot\psi$~(in the plane of Fig.~\ref{param}) and
$c_2=(\sin\psi)/r$~(perpendicular to the plane of Fig.~\ref{param}).
Enforcing the constraints by Lagrange multipliers, the equilibrium shapes can
be obtained by minimizing the following functional~\cite{seifert91}:
\begin{subequations}
\begin{eqnarray}\nonumber
F^\star[r(s),\psi(s), s_1] &=& 
\pi r(0)^2 \left(\Sigma - W\right)\\
\label{freeenergy}
&+&
\int_{0}^{s_1}\!{\cal L}(r, \dot r, \psi, \dot \psi, \gamma)\,\, ds,
\end{eqnarray}
where
\begin{eqnarray}
{\cal L} &=&  
    2\pi 
	           r\left[ 
	                  \frac{1}{2} \kappa \left(
                                                \dot\psi + \frac{\sin\psi}{r}
                                          \right)^2
				   + \Sigma + \frac{P}{2} r \sin \psi
                 \right]
          \nonumber\\
                &+& 2\pi\gamma (s) \left(
                                   \dot{r} - \cos \psi
                             \right).
\label{freeenergy2}
\end{eqnarray}
\end{subequations}
Here $\psi(s)$ and $r(s)$ are regular functions satisfying
\begin{equation}
\psi(0)=0,\quad\psi(s_1)=\pi, \quad\text{and}\quad r(s_1)=0,
\label{bords}
\end{equation}
while $r(0)\equiv L$ and $s_1$ are arbitrary. The above conditions are
necessary for the vesicle's shape to be closed and in order to avoid
discontinuities of the membrane's normal. The parameters $\Sigma$ and $P$ are
the Lagrange multipliers associated with the area and volume constraints,
respectively. The function $\gamma(s)$ is a field of Lagrange multipliers
enforcing the condition $\dot r = \cos\psi$ for every $s$: this allows to
treat $r(s)$ and $\psi(s)$ as independent functions in the first variation of
$F^\star$ while ensuring that $r(s)$ and $\psi(s)$ effectively parameterize
the same shape.

The first variation of $F^\star$ can be written as
\begin{subequations}
\begin{eqnarray}
\delta F^\star=\int_0^{s_1}\!ds&&\!\!\!\!\left[
\left(
\frac{\partial {\cal L}}{\partial \psi} 
-\frac{d}{ds} \frac{\partial {\cal L}}{\partial \dot \psi}
\right)\delta\psi(s)\right.\nonumber\\
&&+\left.
\left(
\frac{\partial {\cal L}}{\partial r} 
-\frac{d}{ds} \frac{\partial {\cal L}}{\partial \dot r}
\right)\delta r(s)
\right]
+\delta F^\star_b,\qquad
\end{eqnarray}
with
\begin{eqnarray}
\delta F^\star_b&=&
\frac{\partial{\cal L}}{\partial\dot\psi}(s_1)\,\delta\psi(s_1)-
\frac{\partial{\cal L}}{\partial\dot\psi}(0)\,\delta\psi(0)\nonumber\\
&+&
\frac{\partial{\cal L}}{\partial\dot r}(s_1)\,\delta r(s_1)-
\frac{\partial{\cal L}}{\partial\dot r}(0)\,\delta r(0)\nonumber\\
&+&
2\pi\left(\Sigma-W\right)\,r(0)\,\delta r(0)+{\cal L}(s_1)\,\delta s_1.
\label{deltaFb}
\end{eqnarray}
\end{subequations}
The membrane's equilibrium equations are obtained by setting to zero the
coefficients of $\delta\psi(s)$ and $\delta r(s)$ in $\delta F^\star$:
\begin{subequations}
\label{eqeq}
\begin{eqnarray}
\label{eqeq1}
0&=&\ddot\psi-\frac{\gamma \sin\psi}{\kappa r} 
              - \frac {P r \cos\psi}{2 \kappa}
              + \frac{\dot\psi \cos\psi}{r}
              - \frac{\sin2\psi}{2 r^2},\qquad\\
0&=&\dot\gamma-\frac{1}{2} \kappa\left( \dot\psi^2 
                                 - \frac{\sin^2\psi}{r^2} 
                           \right)
               - \Sigma - P r \sin\psi.
\label{eqeq2}
\end{eqnarray}
\end{subequations}
The constraint $\dot r = \cos\psi$, which determines the Lagrange field
$\gamma(s)$, constitutes actually a supplementary differential equation to be
fulfilled.  It is worth noticing that it can be obtained by varying
$F^\star$ with respect to $\gamma(s)$, since
\begin{equation}
\frac{\partial {\cal L}}{\partial \gamma}
- \frac{d}{ds} \left(\frac{\partial {\cal L}}{\partial \dot \gamma}
\right) 
= 0 \quad\Leftrightarrow\quad \dot r = \cos\psi.
\label{eqeq3}
\end{equation}
By analogy with Lagrangian mechanics, $s$ playing the role of time, there
exists therefore a conserved Hamiltonian~${\cal H}$, given
by~\cite{seifert91}
\begin{eqnarray}\nonumber
{\cal H} &=& {\cal L} - \dot \psi \frac{\partial{\cal L}}{\partial \dot \psi}
        - \dot r \frac{\partial{\cal L}}{\partial \dot r}
	   - \dot \gamma \frac{\partial{\cal L}}{\partial \dot \gamma}\\
&=&  2\pi r\left[  
	                  \frac{1}{2} \kappa \left(
                                                \dot\psi^2  
                                                  - 
                                                        \frac{\sin^2\psi}
                                                             {r^2}
                                          \right)
				             \right. \nonumber\\
&\,&\,\,\,\,\,\,\,\,\,\,\,\,\,
          \Bigg.       
                + \frac{\gamma}{r} \cos\psi
                - \Sigma - \frac{P}{2} r \sin \psi
              		\Bigg]\label{Hexpr}.
\label{frstintgr}
\end{eqnarray}
For an equilibrium solution, ${\cal H}$ does not depend on $s$.

The boundary equilibrium equations are obtained by setting to zero the
variation $\delta F^\star_b$ in Eq.~(\ref{deltaFb}). Taking into account
Eqs.~(\ref{bords}) yields $\delta\psi(0)=0$,
$\delta\psi(s_1)=-\dot\psi(s_1)\,\delta s_1$ and $\delta r(s_1)=-\dot
r(s_1)\,\delta s_1$. Therefore
\begin{equation}
\delta F^\star_b={\cal H}(s_1)\,\delta s_1
+2\pi\left[\left(\Sigma-W\right)r(0)-\gamma(0)\right]\delta r(0).
\end{equation}
Since $\delta s_1$ and $\delta r(0)$ are independent, we obtain
\begin{subequations}
\begin{eqnarray}\label{grandH}
{\cal H}(s_1)&=&0,\\
(\Sigma-W)r(0)&=&\gamma(0).
\end{eqnarray}
\end{subequations}
Since ${\cal H}(s)$ is a constant, Eq.~(\ref{grandH}) implies ${\cal
H}(0)\equiv{\cal H}=0$. This yields
$\frac{1}{2}\kappa\dot\psi^2(0)+\gamma(0)/r(0)-\Sigma=0$. Hence the above
conditions can be rewritten as 
\begin{subequations}
\begin{eqnarray}
{\cal H}&=&0,\\
\label{curvdec}
\frac{1}{2}\kappa\,\dot\psi^2(0)&=&W.
\end{eqnarray}
\end{subequations}
Note that Eq.~(\ref{curvdec}) is the familiar curvature boundary
condition for adhering membranes and thin elastic
plates~\cite{seifert90,landau_book_elasticity}. Together with
Eqs.~(\ref{bords}), these equations form the boundary conditions of the
problem. Note that we have $5$ boundary conditions for a fourth-order
system since $s_1$ is also an unknown.

\subsection{First-order corrections to the limit $W$ infinite}

In order to compute the first-order corrections to the limit $W$
infinite, we shall determine the shape of the contact angle region in
the case of strong adhesion. To this aim, we first integrate once the
membrane equilibrium equations by replacing Eq.~(\ref{eqeq2}) by the
integral condition
${\cal H}=0$:
\begin{subequations}\label{eqraccord}
\begin{eqnarray} 
\label{eqraccord1}
\ddot\psi &=& \frac{\gamma \sin\psi}{\kappa r} 
              + \frac {P r \cos\psi}{2 \kappa}
              - \frac{\dot\psi \cos\psi}{r}
              + \frac{\sin \left( 2 \psi \right)}{2 r^2},\quad\\
\label{eqraccord3}
\gamma&=&\frac{r}{\cos\psi}\left[
\Sigma+\frac{P}{2}r\sin\psi-
\frac{1}{2}\kappa\left(\dot\psi^2-\frac{\sin^2\psi}{r^2}\right)
\right],\qquad\\
\label{eqraccord2}
\dot r &=& \cos\psi.
\end{eqnarray}
\end{subequations}

\subsubsection{Shape of the contact angle region}
\label{CAshape}

\begin{figure}
\centerline{\includegraphics[width=8cm]{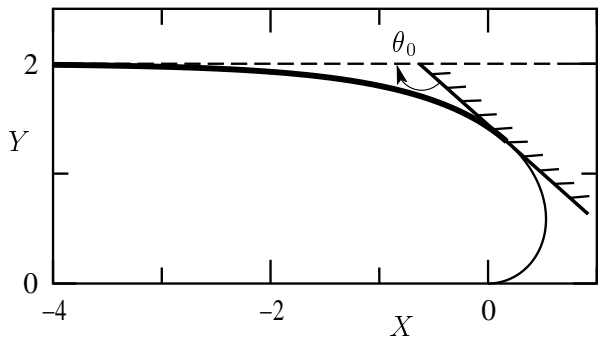}}
\caption{Dimensionless universal shape of the contact angle region. For a
given contact angle $\theta_0$, the actual shape, in units
of~$\ell=\sqrt{\kappa(1+\cos\theta_0)/W}$, is the part of the curve comprised
between the horizontal asymptote and the substrate, the latter being the
tangent to the curve oriented at the angle $\theta_0$ with respect to the
asymptote.}
\label{univshape}
\end{figure}

The equilibrium problem embodied in Eqs.~(\ref{eqraccord})
cannot be solved analytically. As evidenced by the boundary
condition~(\ref{curvdec}), the width of the contact angle region (see
Fig.~\ref{schema_general}) is of order $\sqrt{\kappa/W}$; hence the condition
of strong adhesion can be expressed as
\begin{equation}\label{strgadh2}
\epsilon=\frac{1}{L}\sqrt{\frac{\kappa}{W}}\ll1,
\end{equation}
where $L=r(0)$ is the adhesion disc's radius. This condition
refines~(\ref{strgadhcond}). We therefore start with the estimates: 
\begin{subequations}
\label{ode1}
\begin{eqnarray}
\dot\psi(s)&=&\sqrt{\frac{W}{\kappa}}\times\mathcal{O}(1),\\
r(s)&=&L_0\left[1+o(1)\right],\\
\Sigma&=&\Sigma_0\left[1+o(1)\right],\\
P&=&P_0\left[1+o(1)\right],
\end{eqnarray}
\end{subequations}
where $o(1)$ indicates terms that tend to zero with $\epsilon$ and ${\cal
O}(1)$ indicates terms of order unity. It follows that in
Eq.~(\ref{eqraccord3}) all the terms in the brackets are equal to
$W\times{\cal O}(1)$ except the last one which equals $W\times{\cal
O}(\epsilon^2)$. We therefore neglect it, which amounts to neglecting the
orthoradial principal curvature $(\sin\psi)/r$; thus Eq.~(\ref{eqraccord3})
can be rewritten as
\begin{equation}
\gamma(s)=\frac{L_0}{\cos\psi}\left[
\Sigma_0+\frac{P_0}{2}L_0\sin\psi-
\frac{1}{2}\kappa\dot\psi^2\right]\left[1+o(1)\right].
\end{equation} 
Plugging this expression of $\gamma(s)$ into Eq.~(\ref{eqraccord1}) and using
Eqs.~(\ref{ode1}), we obtain
\begin{eqnarray}
\ddot\psi&=&\left[
\frac{\sin\psi}{\kappa\cos\psi}
\left(
-\frac{1}{2}\kappa\dot\psi^2+\Sigma_0+\frac{P_0L_0}{2}\sin\psi
\right)\right.\nonumber\\
&+&\left.
\frac{P_0L_0\cos\psi}{2\kappa}-\frac{\dot\psi\cos\psi}{L_0}
+\frac{\sin2\psi}{2L_0^2}\right]\left[1+o(1)\right].\quad
\end{eqnarray}
All the terms in this equation are equal to $W\kappa^{-1}\times
\mathcal{O}(1)$, except the last two term which are equal to
$W\kappa^{-1}\times\mathcal{O}(\epsilon)$ and
$W\kappa^{-1}\times\mathcal{O}(\epsilon^2)$, respectively. Using the
expressions of the zeroth-order Lagrange multipliers~(\ref{youngrelat})
and~(\ref{Laplacerelat}), we obtain finally
\begin{equation}\label{psisimple}
\ddot \psi = \left(- \frac{1}{2} \dot\psi^2 \tan\psi
             + \frac{W}{\kappa}
              \frac{\sin\psi - \sin\theta_0}{\left(
                                            1 + \cos\theta_0
						\right)\cos\psi}\right)
\left[1+o(1)\right].
\end{equation}
Neglecting the $o(1)$ term provides us with a simplified equation describing
the contact angle region in the regime of strong adhesion.  This equation can
easily be integrated once by introducing the intermediate variable
$\dot\psi^2/(2\cos\psi)$ and using the boundary condition~(\ref{curvdec}):
\begin{equation}\label{zresolu}
\dot\psi^2 = \frac{2 W}{\kappa}  \frac{1 + \cos\left(\theta_0 + \psi\right)}
                                    {1 + \cos\theta_0}.
\end{equation} 
Its solution is
\begin{equation}\label{sollasido}
\psi(s)=4\arctan\left[\tanh\left(s\sqrt{\frac{W}
{4\kappa\left(1+\cos\theta_0\right)}}\right)\right]-\theta_0,
\end{equation}
where we have shifted the arc-length $s$ by a constant, the detachment
point still corresponding to $\psi=0$. Since the radius $L$ of the adhesion disc
has disappeared, the problem has actually become two-dimensional, as if the
rim of the contact angle region were translationally invariant. This implies
that in the present regime of strong but finite adhesion, the size of the
vesicle has no influence on the shape of the contact angle region. Yet, the
constraint on the reduced volume keeps an influence since it determines
$\theta_0$.

Scaling lengths to $\ell=\sqrt{\kappa(1+\cos\theta_0)/W}$, and introducing a
normalized frame $(X,Y)$ rotated of an angle $\theta_0$ with respect to the
frame $(r,z)$, the shape of the contact angle region assumes the universal
expression:
\begin{subequations}
\begin{eqnarray}
X(S) &=& 2\tanh S - S,\\
Y(S) &=& 2\left[1- \left(\cosh S\right)^{-1}\right],
\end{eqnarray}
\end{subequations}
where $S=s/\ell$ is the normalized arc-length. For a given contact
angle~$\theta_0$, the actual shape of the contact angle region is obtained by
putting the substrate tangent to this shape, at the angle $\theta_0$ with
respect to the horizontal asymptote of the curve, and then rescaling lengths
with respect to~$\ell$, as shown in Fig.~\ref{univshape}. This shape is the
same as that found in~\cite{servuss89}, which was established using an open
membrane description and by imposing the asymptotic direction of the membrane
at the angle $\theta_0$ through an externally imposed tension. 

\subsubsection{Contact angle extrapolation length}
\label{extrapol}

A useful characteristic of the contact angle region, used in RICM experiments
in order to determine the ratio $W/\kappa$~\cite{nardi98,guttenberg00}, is the
\textit{extrapolation length}~$\lambda_1$ (see Fig.~\ref{schema_general}).
From the above calculation, valid in the regime of strong adhesion, we deduce
\begin{equation}
\label{lambda1}
\lambda_1\simeq\int_0^\infty\!\cos\psi\,ds-
\int_0^\infty\!\frac{\sin\psi}{\tan\theta_0}ds
=\sqrt{\frac{2 \kappa}{W}} 
\cot\frac{\theta_0}{2}.
\end{equation}
This expression holds even for deflated vesicles and agrees with the
expression previously obtained in Ref.~\cite{guttenberg00} for nearly
spherical vesicles ($\pi-\theta_0\ll1$).

\subsubsection{First-order corrections to the global observables}
\label{order1}

Let us determine, in the regime of strong but finite adhesion, the global
observables characterizing the vesicle's shape: $\theta$, $R$, $L$, $H$ (see
Sec.~\ref{presentgen} for their definitions). To this purpose, we match the
contact angle region to the rest of the vesicle.  This is done by expressing
the area and volume constraints:
\begin{subequations}
\label{eqAV}
\begin{eqnarray}
\label{eqA}
A&=&A_\mathrm{cap}-\delta A,\\
\label{eqV}
V&=&V_\mathrm{cap}-\delta V,
\end{eqnarray}
\end{subequations}
where $A_\mathrm{cap}=\pi R^2[2(1-\cos\theta)+\sin^2\theta]$ is the area of
the spherical cap osculatory to the vesicle plus the area of its bounding
disc, $V_\mathrm{cap}=\frac{1}{3}\pi
R^3[2(1-\cos\theta)-\sin^2\theta\cos\theta]$ is the volume enclosed by this
spherical cap, and $A$ and $V$ are the actual vesicle's area and volume,
respectively.

In the regime of strong adhesion, $\delta A$ can be evaluated from the results
of Sec.~\ref{CAshape} by calculating the difference between the area
associated with the approximate contact angle shape given by
Eq.~(\ref{sollasido}) and that associated with its asymptote:
\begin{eqnarray}\label{areacost1}
\delta A &\simeq& 2 \pi L_0 \int_0^\infty\!ds
-2\pi L_0\left(
\lambda_1+\int_0^\infty\!\frac{\sin\psi}{\sin\theta_0}ds
\right)\nonumber\\
\label{areacost2}
         &=&
4\pi\left(
\cos\frac{\theta_0}{2}-\cot\frac{\theta_0}{2}\right)
             \sqrt{\frac{2\kappa}{W}}L_0=A\times\mathcal{O}(\epsilon),\quad
\end{eqnarray}
where $L_0=R_0\sin\theta_0$. As for $\delta V\approx\delta A\,\lambda_1$,
it follows that it is equal to $V\times\mathcal{O}(\epsilon^2)$ since
$\lambda_1=L\times\mathcal{O}(\epsilon)$ [see Eq.~(\ref{lambda1})]. Note also
that since in the limit $W\to\infty$ the vesicle's shape is actually a
spherical cap, we have $A=A_\mathrm{cap}(R_0,\theta_0)$ and
$V=V_\mathrm{cap}(R_0,\theta_0)$. 

Setting $\theta=\theta_0+\delta\theta$ and $R=R_0+\delta R$, we
obtain the first-order corrections $\delta\theta$ and $\delta R$ by solving
the system~(\ref{eqAV}) to first order in $\epsilon$. This yields
\begin{subequations}
\label{dtheta-dR}
\begin{eqnarray}
\label{deltatheta}
\delta\theta &=& 
\frac{2\left(\sin\displaystyle\frac{\theta_0}{2}-1\right)
\left(2 +\cos\theta_0\right)}{R_0\sin \theta_0}\,
\sqrt{\frac{2\kappa}{W}}
+\mathcal{O}(\epsilon^2),\qquad\\
\label{deltaR}
\delta R &=& 
\frac{2\left(1-\sin\displaystyle\frac{\theta_0}{2}\right)\sin^2\theta_0}
{\left( 1 - \cos \theta_0 \right)^2}
\sqrt{\frac{2\kappa}{W}}+\mathcal{O}(R_0\epsilon^2).
\end{eqnarray}
\end{subequations}
These results show that, in order to compensate the area cost $\delta A$ of
the contact angle region, the vesicle's shape flattens ($\delta R<0$) with
respect to the asymptotic case of infinite adhesion.

\begin{figure}
\centerline{\includegraphics[width=8cm]{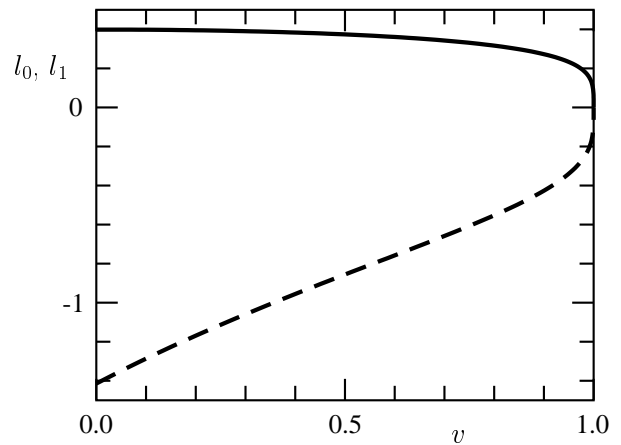}}
\caption{Coefficients $l_0$ (solid line) and $l_1$ (dashed line) of the
expansion~(\protect\ref{Ldev}) of the radius $L$ of the adhesion disc,
as a function of the reduced volume $v$ of the vesicle.}
\label{Lplot}
\end{figure}

We are now able to determine the fist-order corrections to $L$, the radius of
the adhesion disc, and to $H$, the height of the vesicle. Since the
intersection between the substrate and the osculatory spherical cap, described
by $(R, \theta)$, is a circle of radius $R \sin \theta$, we have $L \simeq R
\sin \theta - \lambda_1$.  Using Eqs.~(\ref{dtheta-dR}) and~(\ref{lambda1}),
$L$ can be written in the dimensionless form:
\begin{subequations}
\label{LLdev}
\begin{equation}\label{Ldev}
\frac{L}{\sqrt{A}} = l_0 + l_1 \sqrt{\frac{\kappa}{W A}} 
                         + {\cal O}(\frac{\kappa}{W A}),
\end{equation}
with,
\begin{eqnarray}\label{l0dev}
l_0 &=& \frac{L_0}{\sqrt{A}} 
    = \sqrt{\frac{1+\cos\theta_0}{\pi\left(3+\cos\theta_0\right)}},\\
\label{l1dev}
l_1 &=& - \sqrt{2}\frac{\cos \displaystyle\frac{\theta_0}{2}}
                    {1 + \sin \displaystyle\frac{\theta_0}{2}}.
\end{eqnarray}
\end{subequations}
Note that $\theta_0$ is linked to the prescribed reduced volume~$v$ of the
vesicle through expression~(\ref{vdetheta}). In Fig.~\ref{Lplot} we have
plotted $l_0$ and $l_1$ as a function of $v$.

\begin{figure}
\centerline{\includegraphics[width=8cm]{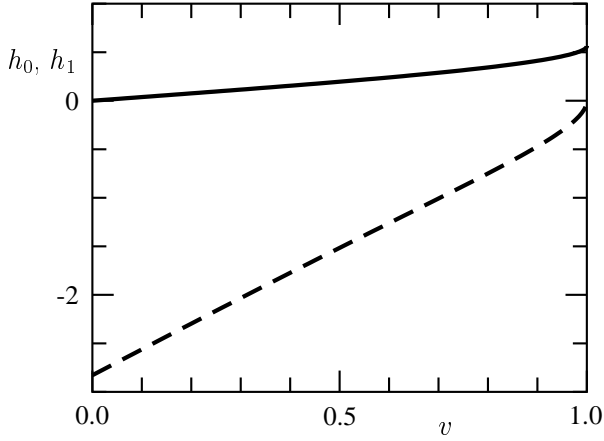}}
\caption{Coefficients $h_0$ (solid line) and $h_1$ (dashed line) of the
expansion~(\protect\ref{Hdev}) of the total height of the vesicle, as a
function of the reduced volume $v$ of the vesicle.}
\label{Hplot}
\end{figure}

As for $H$, since the osculatory spherical cap is tangent to the top of the
vesicle, we have simply $H=R(1 - \cos \theta)$. Using Eqs.~(\ref{dtheta-dR})
we obtain
\begin{subequations}
\begin{equation}\label{Hdev}
\frac{H}{\sqrt{A}} = h_0 + h_1 \sqrt{\frac{\kappa}{W A}} 
                         + {\cal O}(\frac{\kappa}{W A}),
\end{equation}
with,
\begin{eqnarray}\label{h0dev}
h_0 &=& \frac{H_0}{\sqrt{A}} 
=\sqrt{\frac{1-\cos\theta_0}{\pi\left(3+\cos\theta_0\right)}},\\
\label{h1dev}
h_1 &=& - 2 \sqrt{2} \left( 1 - \sin \frac{\theta_0}{2}\right).
\end{eqnarray}
\end{subequations}
The plots of $h_0$ and $h_1$ as a function of $v$ are shown in
Fig.~\ref{Hplot}.  Note that $h_0$ and $l_0$ stem from simple geometrical
considerations, while $h_1$ and $l_1$ originate from curvature elasticity
effects.  

\subsubsection{Free energy of adhering vesicles}
\label{freeen}

We now turn to the determination of the analytical development of the total
free energy of the vesicle:
\begin{equation}
F=-\pi L^2\,W+F_\mathrm{el},
\end{equation}
where $F_\mathrm{el}$ is the curvature free energy. The latter is the sum of a
contribution $F_\mathrm{el,1}$ arising from the contact-angle region and a
contribution $F_\mathrm{el,2}$ arising from the top spherical cap. Since both
the size and the curvature radius of the contact-angle region are of order
$\sqrt{\kappa/W}$, $F_\mathrm{el,1}$ is of order
$L\sqrt{\kappa/W}\times\kappa(\sqrt{W/\kappa})^2=
WL^2\times\mathcal{O}(\epsilon)$. As for $F_\mathrm{el,2}$, it can be
neglected since it is of order
$\kappa\times\mathcal{O}(1)=WL^2\times\mathcal{O}(\epsilon^2)$, as for a free
vesicle.

\begin{figure}
\centerline{\includegraphics[width=8cm]{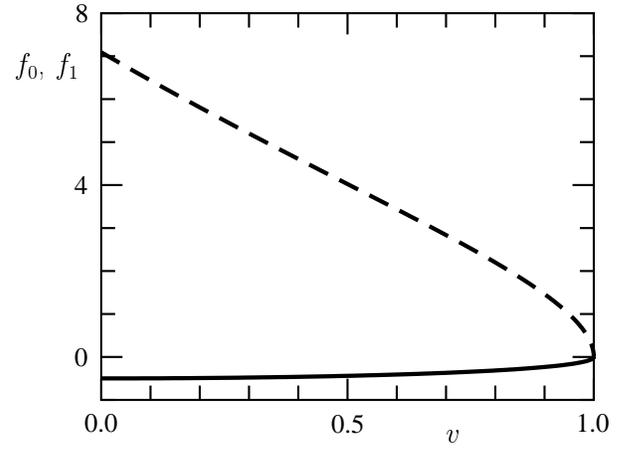}}
\caption{Coefficients $f_0$ (solid line) and $f_1$ (dashed line) of the
expansion~(\protect\ref{Fdev}) of the free energy of the vesicle, as a
function of the reduced volume $v$ of the vesicle.}
\label{Fplot}
\end{figure}

In the strong adhesion regime, the orthoradial curvature $(\sin\psi)/r$ of the
contact-angle region is negligible as justified in Sec.~\ref{CAshape}.
Therefore, using Eq.~(\ref{zresolu}), we obtain
\begin{equation}
F_\mathrm{el} \simeq \pi \kappa L_0 \int_{0}^{\infty}
           \dot{\psi}^2 ds
       = 2\pi L_0 \sqrt{2\kappa W} \frac{1 -
       \sin\displaystyle\frac{\theta_0}{2}}{\cos\displaystyle
       \frac{\theta_0}{2}}.
\label{deltaFel}
\end{equation}
Using the expression of $L$ given by Eq.~(\ref{Ldev}), we finally obtain
\begin{subequations}
\begin{equation}\label{Fdev}
\frac{F}{W A} = f_0 + f_1 \sqrt{\frac{\kappa}{W A}} 
                         + {\cal O}(\frac{\kappa}{W A}),
\end{equation}
with
\begin{eqnarray}\label{f0dev}
f_0 &=&  - \frac{1+\cos\theta_0}{3+\cos \theta_0},\\ 
\label{f1dev}
f_1 &=& 8 \sqrt{\pi} \frac{1 - \sin \displaystyle\frac{\theta_0}{2}}
                          {\sqrt{3 + \cos \theta_0}}.
\end{eqnarray}
\end{subequations}
The plots of $f_0$ and $f_1$ in terms of the reduced volume $v$ are shown in
Fig.~\ref{Fplot}. 

As an application of this result, let us determine the force acting on an
adhering vesicle in the presence of weak adhesion gradients:
haptotaxis~\cite{cantat00}. If the dynamical deformations
during the movement are weak, the shape of the vesicle can be assimilated to
its equilibrium shape on a substrate with a constant adhesion potential $W$
equal to the average of $W$ in the real adhesion disc. The force exerted on
the vesicle is then 
\begin{eqnarray}\nonumber
\mathbf{f} &=& - \frac{\partial F}{\partial W} \nabla W\\
\label{haptotaxdev}
          &=& -\left[ 
                     f_0 A + \frac{1}{2} f_1\sqrt{\frac{\kappa A}{W}} 
                     + {\cal O}\left( \frac{\kappa}{W}\right)
               \right] 
              \nabla W,
\end{eqnarray}
where $\nabla$ is the gradient on the substrate. Since $f_0$ and $f_1$ have
opposite signs, the curvature elasticity decreases the
haptotactic force with respect to the infinite adhesion limit. Moreover,
for a given $\nabla W$ the haptotactic force is not constant
but actually increases with~$W$.

\begin{figure}
\centerline{\includegraphics[width=8cm]{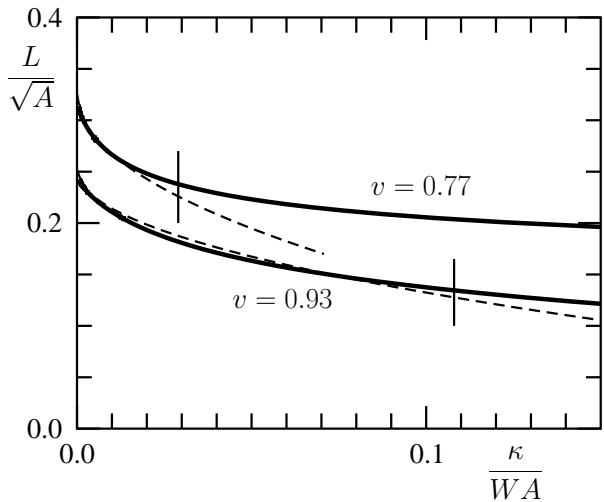}}
\caption{Numerically calculated radius $L$ of the adhesion disc as a
function of $\kappa/(W A)$ (solid lines) for vesicles of reduced volume
$v=0.77$ and $v=0.93$, along with its asymptotic
expansion~(\protect\ref{LLdev}) (dashed lines). The vertical bars
indicate the threshold at which the relative error between the exact
value of $L$ and its analytical estimate reaches~$5\%$.}
\label{LdeW}
\end{figure}

To check the order of magnitude of the haptotactic force, let us
consider a $10\,\mu$m vesicle ($A\simeq10^{-9}$\,m$^2$) with
$\kappa\simeq10^{-19}$\,J, subject to a contact potential varying
uniformly from $W\simeq10^{-4}$\,mJ$/\mathrm{m}^2$ to
$W\simeq10^{-3}$\,mJ$/\mathrm{m}^2$ on a distance $\simeq1$\,mm.
Assuming a reduced volume $v=0.77$ corresponding to
$\theta_0\simeq\pi/2$ [see Eq.~(\ref{vdetheta})], we obtain a force
varying from $0.26\,$pN to $0.29\,$pN ($8\%$ variation).  With a simple
Stokes law, this corresponds to velocities of the order of
$1\,\mathrm{\mu}$m\,s$^{-1}$. Note that in infinite adhesion this
gradient would give rise to a force equal to $0.3\,$pN.

\section{Comparison with the exact numerical results}\label{ordre1lim}

We expect the asymptotic expansions given in Secs.\ \ref{order1}
and~\ref{freeen} to be accurate in the regime of strong adhesion. To
check their validity, we have compared them with the exact values of the
vesicle's observables, obtained by numerically integrating Eqs.\
(\ref{eqeq}) and~(\ref{eqeq3}).  

In order to avoid numerical
instabilities when approaching the axis of revolution ($r=0$), we have
chosen to integrate the equations starting from the top of the vesicle
($s=s_1$, see Fig.~\ref{param}). To this aim, we impose the $4$ initial
conditions:
\begin{subequations}
\begin{eqnarray}\label{bcnum1}
\mathcal{H}(s_1)         &=& 0,\\
\label{bcnum2}
r(s_1)         &=& 0,\\
\label{bcnum3}
\psi (s_1)     &=& 0,\\
\label{bcnum4}
\dot\psi (s_1) &=& c_0,
\end{eqnarray}
\end{subequations}
where $\mathcal{H}(s)$ is the first integral of the equilibrium
equations given by Eq.~(\ref{frstintgr}) and $c_0$ is an arbitrary
initial curvature. The integration proceeds backwards, starting from $s
= s_1$ (the actual value of $s_1$ is arbitrary), and is stopped when
$\psi = \pi$, meaning that the substrate has been reached.  To span more
easily all the values of the dimensionless parameter $W A/\kappa$ for a
given reduced volume $v$, we proceed as follows.  For a given fixed
value of the initial curvature~$c_0$, we vary the Lagrange
multiplier~$\Sigma$ in Eqs.~(\ref{eqeq}) until the solution has the
desired reduced volume.  During this search, the other Lagrange
multiplier, $P$, is fixed to a value (positive for
weakly adhering vesicles and negative for strongly adhering
vesicles) assuring that the size of the vesicle is of order~$1$ in
dimensionless units. Once the correct value of $\Sigma$ has been
obtained, we determine the area of the vesicle and its curvature at the
point $\psi =\pi$, where it touches the substrate. The corresponding
value of $W A/\kappa$ is obtained through the boundary
condition~(\ref{curvdec}). The set of all the solutions for a fixed $v$
and all values of $W A/\kappa$ corresponds to a trajectory in the
($\Sigma, c_0$) plane that has to be reconstructed by varying $c_0$ and
$\Sigma$. Sometimes, a given $c_0$ corresponds to two or more values of
$\Sigma$, which yields different values of $W A/\kappa$ for
the same reduced volume.

\begin{figure}
\centerline{\includegraphics[width=8cm]{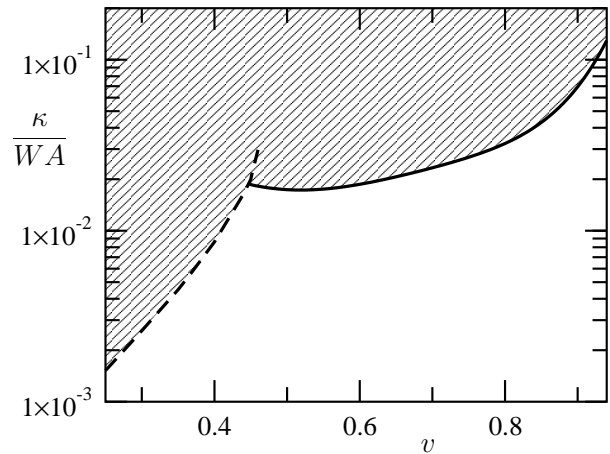}}
\caption{White area: region of the $(v,\kappa/WA)$ plane where the analytical
estimate of the radius $L$ of the adhesion disc differs from its exact
numerical evaluate by less than $5\%$.  Above the dashed line (not fully
shown) the axisymmetric oblate shapes correspond to an unphysical
self-crossing of the membrane.}
\label{ordre1limL}
\end{figure}

To exemplify our results, we show in Fig.~\ref{LdeW} the radius~$L$ of
the adhesion disc, as a function of the reduced inverse adhesion energy
$\kappa/(W A)$, for vesicles of reduced volume $v=0.77$ (or $\theta_0
\simeq \pi/2$) and $v = 0.93$. High adhesion energies correspond to low
values of $\kappa/(W A)$, where our asymptotic formula~(\ref{LLdev})
closely fits the exact numerical results. The vertical bars indicate the
threshold above which the error associated with the analytical
approximation is larger than $5\%$. At this threshold, $L$ differs
nonetheless from its infinite adhesion limit $L_0$ by more than $30 \%$:
significant deviations from the infinite adhesion limit are therefore
predicted with a good precision by the asymptotic formula~(\ref{LLdev}).  

\begin{figure}
\centerline{\includegraphics[width=8cm]{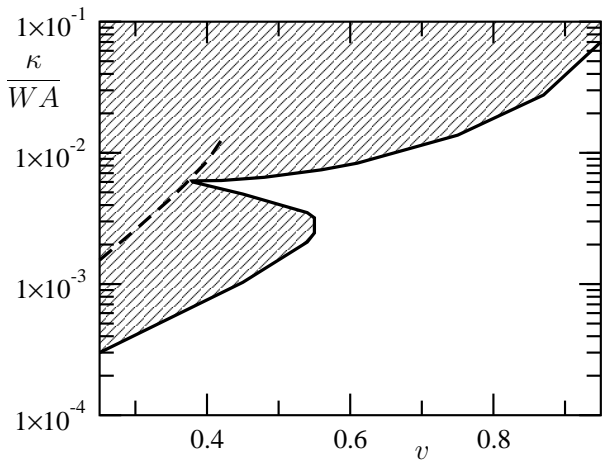}}
\caption{Same as Fig.~\protect\ref{ordre1limL} but for the total height
$H$ of the vesicle.}
\label{ordre1limH}
\end{figure}

\begin{figure}
\centerline{\includegraphics[width=8cm]{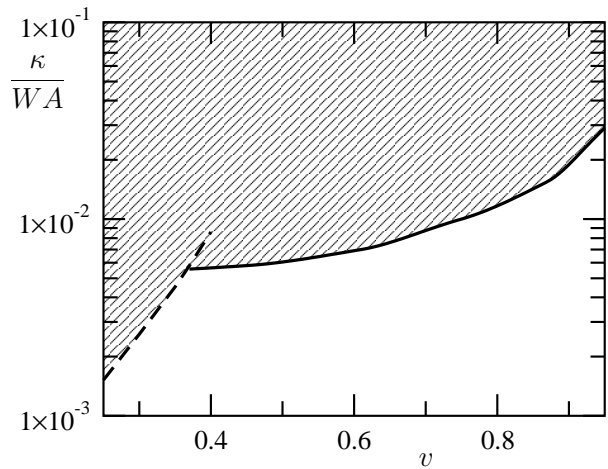}}
\caption{Same as Fig.~\protect\ref{ordre1limL} but for the derivative
$dF/dW$ of the free energy of the vesicle with respect to the adhesion
energy~$W$.}
\label{ordre1limF}
\end{figure}

For vesicles of reduced volume in the range $0.25\le v\le 0.95$, we have
determined the threshold for $\kappa/(W A)$ at which the relative error between
our analytical approximations and the exact results reaches $5\%$. In
Fig.~\ref{ordre1limL} we show this threshold for the adhesion disc's
radius~$L$, in Fig.~\ref{ordre1limH} for the total vesicle's height~$H$, and
finally in Fig.~\ref{ordre1limF} for the derivative $dF/dW$ of the free
energy with respect to the adhesion energy. The latter quantity is
linked to the haptotactic force~(\ref{haptotaxdev}).

Typically, the $5\%$ threshold occurs for values of $\kappa/(W A)$
comprised between $10^{-2}$ and $10^{-3}$ (see
Figs.~\ref{ordre1limL}--\ref{ordre1limF}).  Let us consider the case of
``giant vesicles'' since they are optically observable (typical size
$\simeq 10$--$100\, \mu \mathrm{m}$). Supposing an area of $\simeq
10^3\, \mu \mathrm{m}^2$ and a bending rigidity $\kappa\simeq
10^{-19}\,$J, the $5\%$ threshold occurs for values of $W$ in the weak
adhesion range $ 10^{-5}$--$10^{-4}\,\mathrm{mJ}/\mathrm{m}^{2}$. Our
analytical estimates seem therefore able to describe the adhesion of
giant vesicles up to the lowest values of $W$ experimentally accessible.
For smaller vesicles, the threshold is more limitative, as it
corresponds to higher adhesion energies $W$. Note also that in the case
of weak adhesion, the picture could be quantitatively different for
vesicles filled with a fluid denser than the outside medium, because of
gravity effects.

\section{Discussion and possible applications}\label{conclusion}

Taking into account the effect of membrane elasticity to
first-order in $\sqrt{\kappa/(W A)}$, we have analytically determined
the global observables characterizing adhering vesicles. Our calculation
is based on the fact that if adhesion prevails over elasticity, most of
the elastic contributions to the free energy are located in the
``contact angle region.'' We have numerically determined the region of
validity of our analytical expansions, in the $(v,\kappa/W A)$
parameter space, corresponding to a $5\%$ maximum error. It turns out
that for ``giant vesicles'' (typical radius $10$--$100\,\mu$m), this
region comprises practically all the accessible adhesion surface
energies~$W$.  Besides, our analytical estimates correctly describe
significant deviations with respect to the infinite adhesion limit. 

We have throughout assumed that the area~$A$ and volume~$V$ of the
vesicle were strictly fixed, while vesicles actually possess small but
finite stretching elasticity and osmotic compressibility. It is easily
shown, however, that a self-consistent choice of the Lagrange multipliers
$\Sigma$ and $P$ yields the same equilibrium solution in the presence
of arbitrary stretching and osmotic potentials. It follows that our
expressions of $\delta\theta$, $\delta R$, $L$ and $H$ remain correct
provided that $A$ and $V$ are the actual area and volume (that now
depend on $W$).

Measurements of the contact potential $W$ are usually performed by RICM
imaging of the contact angle region~\cite{guttenberg00,nardi98}. The value of
$W$ is inferred either from the local curvature $\dot \psi (0)$ through
Eq.~(\ref{curvdec}), or from the extrapolation length $\lambda_1$ through
Eq.~(\ref{lambda1}) (an approximated formula valid for $\pi-\theta_0\ll1$ is
actually used~\cite{guttenberg00}). The precision of the former measurement is
limited by the fact that the vesicle's curvature varies abruptly close to its
detachment point, the exact position of which is always slightly ambiguous.
The extrapolation length measurement relies on the existence of a well defined
asymptote of the vesicle's profile close to the contact angle region: it is
therefore suitable only for the strongest adhesions. The expressions of
$L$ and $H$ found in Sec.~\ref{order1} allow to envisage novel
measurements of $W$, based on \textit{global} characteristics of the adhering
vesicle. To this aim, one needs to know also the vesicle's total area and
volume. They can be either directly determined by imaging a side view of the
vesicle~\cite{abkarian01}, or inferred by osmotically deflating a spherical
vesicle of known radius in a controlled way. The fact that $W$ can be
determined through two independent measurements ($L$ and $H$) allows to better
estimate the experimental errors and to validate the model.  Moreover,
such global measurements are complementary to the above-cited local
ones, since they are more adapted for measuring weaker values of $W$.
The precision of the measurement should increase as $W$ decreases, as
long as one remains inside the authorized zone of Figs. \ref{ordre1limL}
and~\ref{ordre1limH}. In fact, for $W$ too strong, $L$ and $H$ saturate,
while for $W$ too weak the analytical expansions of $L$ and $H$ lose
their validity.  However, as we have seen, the low $W$ limitation is not
relevant for ``giant vesicles''.

Finally, the haptotactic force~(\ref{haptotaxdev}) suggests the possibility to
determine the size of suboptical vesicles by measuring their velocity of
migration on a substrate presenting a controlled adhesion gradient, supposing
a linear viscous friction law. Fitting the evolution of the vesicle's velocity
as a function of $W$ allows to determine the vesicle's area and volume,
provided that the dependence of the friction coefficient on $\kappa/ (W A)$
and $v$ is known. The latter could be determined using giant vesicles of
known area and volume.


\begin{thebibliography}{24}
\expandafter\ifx\csname natexlab\endcsname\relax\def\natexlab#1{#1}\fi
\expandafter\ifx\csname bibnamefont\endcsname\relax
  \def\bibnamefont#1{#1}\fi
\expandafter\ifx\csname bibfnamefont\endcsname\relax
  \def\bibfnamefont#1{#1}\fi
\expandafter\ifx\csname citenamefont\endcsname\relax
  \def\citenamefont#1{#1}\fi
\expandafter\ifx\csname url\endcsname\relax
  \def\url#1{\texttt{#1}}\fi
\expandafter\ifx\csname urlprefix\endcsname\relax\def\urlprefix{URL }\fi
\providecommand{\bibinfo}[2]{#2}
\providecommand{\eprint}[2][]{\url{#2}}

\bibitem[{\citenamefont{Israelachvili}(1991)}]{israelachvili_book}
\bibinfo{author}{\bibfnamefont{J.}~\bibnamefont{Israelachvili}},
  \emph{\bibinfo{title}{Intermolecular \& Surface Forces}}
  (\bibinfo{publisher}{Academic Press}, \bibinfo{address}{New York},
  \bibinfo{year}{1991}).

\bibitem[{\citenamefont{Sternberg et~al.}(1987)\citenamefont{Sternberg,
  Grumpert, Reinhardt, and Gaw\-risch}}]{sternberg87}
\bibinfo{author}{\bibfnamefont{B.}~\bibnamefont{Sternberg}},
  \bibinfo{author}{\bibfnamefont{J.}~\bibnamefont{Grumpert}},
  \bibinfo{author}{\bibfnamefont{G.}~\bibnamefont{Reinhardt}},
  \bibnamefont{and}
  \bibinfo{author}{\bibfnamefont{K.}~\bibnamefont{Gaw\-risch}},
  \bibinfo{journal}{Biochim. Biophys. Acta} \textbf{\bibinfo{volume}{898}},
  \bibinfo{pages}{223} (\bibinfo{year}{1987}).

\bibitem[{\citenamefont{Guedeau-Boudeville
  et~al.}(1995)\citenamefont{Guedeau-Boudeville, Jullien, and
  di~Me\-glio}}]{guedeau-boudeville95}
\bibinfo{author}{\bibfnamefont{M.-A.} \bibnamefont{Guedeau-Boudeville}},
  \bibinfo{author}{\bibfnamefont{L.}~\bibnamefont{Jullien}}, \bibnamefont{and}
  \bibinfo{author}{\bibfnamefont{J.-M.} \bibnamefont{di~Me\-glio}},
  \bibinfo{journal}{Proc. Natl. Acad. Sci. USA} \textbf{\bibinfo{volume}{92}},
  \bibinfo{pages}{9590} (\bibinfo{year}{1995}).

\bibitem[{\citenamefont{Bernard et~al.}(2000)\citenamefont{Bernard,
  Guedeau-Boudeville, Jullien, and di~Meglio}}]{bernard00}
\bibinfo{author}{\bibfnamefont{A.-L.} \bibnamefont{Bernard}},
  \bibinfo{author}{\bibfnamefont{M.-A.} \bibnamefont{Guedeau-Boudeville}},
  \bibinfo{author}{\bibfnamefont{L.}~\bibnamefont{Jullien}}, \bibnamefont{and}
  \bibinfo{author}{\bibfnamefont{J.-M.} \bibnamefont{di~Meglio}},
  \bibinfo{journal}{Langmuir} \textbf{\bibinfo{volume}{16}},
  \bibinfo{pages}{6809} (\bibinfo{year}{2000}).

\bibitem[{\citenamefont{Keller et~al.}(2000)\citenamefont{Keller,
  Glasm{\"a}star, Zhdanov, and Ka\-semo}}]{keller00}
\bibinfo{author}{\bibfnamefont{C.~A.} \bibnamefont{Keller}},
  \bibinfo{author}{\bibfnamefont{K.}~\bibnamefont{Glasm{\"a}star}},
  \bibinfo{author}{\bibfnamefont{V.~P.} \bibnamefont{Zhdanov}},
  \bibnamefont{and} \bibinfo{author}{\bibfnamefont{B.}~\bibnamefont{Ka\-semo}},
  \bibinfo{journal}{Phys. Rev. Lett.} \textbf{\bibinfo{volume}{84}},
  \bibinfo{pages}{5443} (\bibinfo{year}{2000}).

\bibitem[{\citenamefont{Sackmann}(1996)}]{sackmann96}
\bibinfo{author}{\bibfnamefont{E.}~\bibnamefont{Sackmann}},
  \bibinfo{journal}{Science} \textbf{\bibinfo{volume}{271}},
  \bibinfo{pages}{43} (\bibinfo{year}{1996}).

\bibitem[{\citenamefont{Bruinsma et~al.}(2000)\citenamefont{Bruinsma, Berisch,
  and Sackmann}}]{bruinsma00}
\bibinfo{author}{\bibfnamefont{R.}~\bibnamefont{Bruinsma}},
  \bibinfo{author}{\bibfnamefont{A.}~\bibnamefont{Berisch}}, \bibnamefont{and}
  \bibinfo{author}{\bibfnamefont{E.}~\bibnamefont{Sackmann}},
  \bibinfo{journal}{Phys. Rev. E} \textbf{\bibinfo{volume}{61}},
  \bibinfo{pages}{4253} (\bibinfo{year}{2000}).

\bibitem[{\citenamefont{Seifert}(1997)}]{seifert97}
\bibinfo{author}{\bibfnamefont{U.}~\bibnamefont{Seifert}},
  \bibinfo{journal}{Advances in Physics} \textbf{\bibinfo{volume}{46}},
  \bibinfo{pages}{13} (\bibinfo{year}{1997}).

\bibitem[{\citenamefont{R{\"a}dler et~al.}(1995)\citenamefont{R{\"a}dler,
  Feder, Strey, and Sackmann}}]{radler95}
\bibinfo{author}{\bibfnamefont{J.~O.} \bibnamefont{R{\"a}dler}},
  \bibinfo{author}{\bibfnamefont{T.~J.} \bibnamefont{Feder}},
  \bibinfo{author}{\bibfnamefont{H.~H.} \bibnamefont{Strey}}, \bibnamefont{and}
  \bibinfo{author}{\bibfnamefont{E.}~\bibnamefont{Sackmann}},
  \bibinfo{journal}{Phys. Rev. E} \textbf{\bibinfo{volume}{51}},
  \bibinfo{pages}{4526} (\bibinfo{year}{1995}).

\bibitem[{\citenamefont{Helfrich}(1973)}]{helfrich73}
\bibinfo{author}{\bibfnamefont{H.}~\bibnamefont{Helfrich}},
  \bibinfo{journal}{Z. Naturforsch.} \textbf{\bibinfo{volume}{28C}},
  \bibinfo{pages}{693} (\bibinfo{year}{1973}).

\bibitem[{\citenamefont{Lee et~al.}(2001)\citenamefont{Lee, Lin, and
  Wang}}]{lee01}
\bibinfo{author}{\bibfnamefont{C.-H.} \bibnamefont{Lee}},
  \bibinfo{author}{\bibfnamefont{W.-C.} \bibnamefont{Lin}}, \bibnamefont{and}
  \bibinfo{author}{\bibfnamefont{J.}~\bibnamefont{Wang}},
  \bibinfo{journal}{Phys. Rev. E} \textbf{\bibinfo{volume}{64}},
  \bibinfo{pages}{020901(R)} (\bibinfo{year}{2001}).

\bibitem[{\citenamefont{Svetina et~al.}(1982)\citenamefont{Svetina,
  Ottova-Lietmannova, and Glaser}}]{svetina82}
\bibinfo{author}{\bibfnamefont{S.}~\bibnamefont{Svetina}},
  \bibinfo{author}{\bibfnamefont{A.}~\bibnamefont{Ottova-Lietmannova}},
  \bibnamefont{and} \bibinfo{author}{\bibfnamefont{R.}~\bibnamefont{Glaser}},
  \bibinfo{journal}{J. Theor. Biol.} \textbf{\bibinfo{volume}{94}},
  \bibinfo{pages}{13} (\bibinfo{year}{1982}).

\bibitem[{\citenamefont{Seifert et~al.}(1992)\citenamefont{Seifert, Miao, and
  D{\" o}bereiner}}]{seifert92}
\bibinfo{author}{\bibfnamefont{U.}~\bibnamefont{Seifert}},
  \bibinfo{author}{\bibfnamefont{L.}~\bibnamefont{Miao}}, \bibnamefont{and}
  \bibinfo{author}{\bibfnamefont{H.~G.} \bibnamefont{D{\" o}bereiner}},
  \bibinfo{journal}{Springer Proceedings in Physics}
  \textbf{\bibinfo{volume}{66}}, \bibinfo{pages}{93} (\bibinfo{year}{1992}).

\bibitem[{\citenamefont{Seifert and Lipowsky}(1990)}]{seifert90}
\bibinfo{author}{\bibfnamefont{U.}~\bibnamefont{Seifert}} \bibnamefont{and}
  \bibinfo{author}{\bibfnamefont{R.}~\bibnamefont{Lipowsky}},
  \bibinfo{journal}{Phys. Rev. A} \textbf{\bibinfo{volume}{42}},
  \bibinfo{pages}{4768} (\bibinfo{year}{1990}).

\bibitem[{\citenamefont{de~Gennes}(1985)}]{deGennes85}
\bibinfo{author}{\bibfnamefont{P.-G.} \bibnamefont{de~Gennes}},
  \bibinfo{journal}{Rev. Mod. Phys.} \textbf{\bibinfo{volume}{57}},
  \bibinfo{pages}{827} (\bibinfo{year}{1985}).

\bibitem[{\citenamefont{Servuss and Helfrich}(1989)}]{servuss89}
\bibinfo{author}{\bibfnamefont{R.~M.} \bibnamefont{Servuss}} \bibnamefont{and}
  \bibinfo{author}{\bibfnamefont{W.}~\bibnamefont{Helfrich}},
  \bibinfo{journal}{J. Phys. France} \textbf{\bibinfo{volume}{50}},
  \bibinfo{pages}{809} (\bibinfo{year}{1989}).

\bibitem[{\citenamefont{Guttenberg et~al.}(2000)\citenamefont{Guttenberg,
  Bausch, Hu, Bruinsma, Moroder, and Sackmann}}]{guttenberg00}
\bibinfo{author}{\bibfnamefont{Z.}~\bibnamefont{Guttenberg}},
  \bibinfo{author}{\bibfnamefont{A.~R.} \bibnamefont{Bausch}},
  \bibinfo{author}{\bibfnamefont{B.}~\bibnamefont{Hu}},
  \bibinfo{author}{\bibfnamefont{R.}~\bibnamefont{Bruinsma}},
  \bibinfo{author}{\bibfnamefont{L.}~\bibnamefont{Moroder}}, \bibnamefont{and}
  \bibinfo{author}{\bibfnamefont{E.}~\bibnamefont{Sackmann}},
  \bibinfo{journal}{Langmuir} \textbf{\bibinfo{volume}{16}},
  \bibinfo{pages}{8984} (\bibinfo{year}{2000}).

\bibitem[{\citenamefont{Landau and Lifchitz}(1967)}]{landau_book_elasticity}
\bibinfo{author}{\bibfnamefont{L.}~\bibnamefont{Landau}} \bibnamefont{and}
  \bibinfo{author}{\bibfnamefont{E.}~\bibnamefont{Lifchitz}},
  \emph{\bibinfo{title}{Th\'eorie de l'\'Elasticit\'e}}
  (\bibinfo{publisher}{Mir}, \bibinfo{address}{Moscou}, \bibinfo{year}{1967}).

\bibitem[{\citenamefont{Rosso and Virga}(1999)}]{rosso99}
\bibinfo{author}{\bibfnamefont{R.}~\bibnamefont{Rosso}} \bibnamefont{and}
  \bibinfo{author}{\bibfnamefont{E.~G.} \bibnamefont{Virga}},
  \bibinfo{journal}{Proc. R. Soc. Lond. A} \textbf{\bibinfo{volume}{455}},
  \bibinfo{pages}{4145} (\bibinfo{year}{1999}).

\bibitem[{\citenamefont{Nardi et~al.}(1998)\citenamefont{Nardi, Bruinsma, and
  Sackmann}}]{nardi98}
\bibinfo{author}{\bibfnamefont{J.}~\bibnamefont{Nardi}},
  \bibinfo{author}{\bibfnamefont{R.}~\bibnamefont{Bruinsma}}, \bibnamefont{and}
  \bibinfo{author}{\bibfnamefont{E.}~\bibnamefont{Sackmann}},
  \bibinfo{journal}{Phys. Rev. E} \textbf{\bibinfo{volume}{58}},
  \bibinfo{pages}{6340} (\bibinfo{year}{1998}).

\bibitem[{\citenamefont{Cantat et~al.}(2000)\citenamefont{Cantat, Misbah, and
  Saito}}]{cantat00}
\bibinfo{author}{\bibfnamefont{I.}~\bibnamefont{Cantat}},
  \bibinfo{author}{\bibfnamefont{C.}~\bibnamefont{Misbah}}, \bibnamefont{and}
  \bibinfo{author}{\bibfnamefont{Y.}~\bibnamefont{Saito}},
  \bibinfo{journal}{Eur. Phys. J. E} \textbf{\bibinfo{volume}{3}},
  \bibinfo{pages}{403} (\bibinfo{year}{2000}).

\bibitem[{\citenamefont{Struik}(1961)}]{struik_book}
\bibinfo{author}{\bibfnamefont{D.~J.} \bibnamefont{Struik}},
  \emph{\bibinfo{title}{Lectures on Classical Differential Geometry}}
  (\bibinfo{publisher}{Dover Publications}, \bibinfo{address}{New York},
  \bibinfo{year}{1961}).

\bibitem[{\citenamefont{Seifert et~al.}(1991)\citenamefont{Seifert, Berndl, and
  Lipowsky}}]{seifert91}
\bibinfo{author}{\bibfnamefont{U.}~\bibnamefont{Seifert}},
  \bibinfo{author}{\bibfnamefont{K.}~\bibnamefont{Berndl}}, \bibnamefont{and}
  \bibinfo{author}{\bibfnamefont{R.}~\bibnamefont{Lipowsky}},
  \bibinfo{journal}{Phys. Rev. A} \textbf{\bibinfo{volume}{44}},
  \bibinfo{pages}{1182} (\bibinfo{year}{1991}).

\bibitem[{\citenamefont{Abkarian et~al.}(2001)\citenamefont{Abkarian, Lartigue,
  and Viallat}}]{abkarian01}
\bibinfo{author}{\bibfnamefont{M.}~\bibnamefont{Abkarian}},
  \bibinfo{author}{\bibfnamefont{C.}~\bibnamefont{Lartigue}}, \bibnamefont{and}
  \bibinfo{author}{\bibfnamefont{A.}~\bibnamefont{Viallat}},
  \bibinfo{journal}{Phys. Rev. E} \textbf{\bibinfo{volume}{63}},
  \bibinfo{pages}{041906} (\bibinfo{year}{2001}).

\end{thebibliography}
\end{document}